\documentclass[12pt]{article}
\usepackage{authblk}
\usepackage{graphicx,psfrag,epsf}
\usepackage{enumerate}
\usepackage[utf8]{inputenc}
\usepackage{import}
\usepackage{multicol}
\usepackage{enumitem}
\usepackage{verbatim}
\usepackage{amsmath, amscd, amssymb, amsthm, xspace, graphicx}

\newtheorem*{theorem*}{Theorem}

\newtheorem*{prop*}{Proposition}

\newtheorem*{corollary*}{Corollary}

\newtheorem*{lemma*}{Lemma}
\usepackage[all]{xy}
\usepackage{subcaption}
\usepackage{float}
\usepackage{mathtools} 
\usepackage{listings}
\usepackage{comment}
\usepackage{color}
\usepackage{titlesec}
\usepackage{makecell}
\usepackage{multirow}
\usepackage[
backend=biber,
sorting=none
]{biblatex}
\AtEveryBibitem{%
  \clearfield{volume}%
  \clearfield{number}
  \clearfield{pages}
  \clearfield{issn}
  \clearfield{issue}
  \clearfield{month}
  \clearfield{eprint}
  \clearfield{isbn}}
\usepackage[toc,page]{appendix}
\newcommand{\Expect}[1]{\mathbb{E}\left[ #1 \right]}

\newcommand{\diffd}{\textnormal{d}}

\newcommand{\R}{\mathbb{R}}
\newcommand{\intR}{\int_{\R}}

\newcommand{\Pois}[1]{\mathrm{Poisson}\left(#1\right)}

\newcommand{\rlang}{\texttt{R}}
\newcommand{\GP}[2]{\mathcal{GP}\left(#1,#2\right)}
\newcommand{\tfo}[1]{{}_2F_1\left(#1\right)}
\numberwithin{equation}{section}
\numberwithin{figure}{section}
\numberwithin{table}{section}

\title{\texttt{PoissonRatioUQ}: An R package for band ratio uncertainty quantification.}
\author[1]{Matthew LeDuc\thanks{matthew.leduc@colorado.edu}}
\author[2,1]{Tomoko Matsuo}
 \affil[1]{Department of Applied Mathematics, University of Colorado Boulder}
 \affil[2]{Ann and H.J. Smead Department of Aerospace Engineering Sciences, University of Colorado Boulder}
\date{\today}

\begin{document}

\maketitle

\begin{abstract}
	We introduce an \texttt{R} package for Bayesian modeling and uncertainty quantification for problems involving count ratios. The modeling relies on the assumption that the quantity of interest is the ratio of Poisson means rather than the ratio of counts. We provide multiple different options for retrieval of this quantity for problems with and without spatial information included. Some added capability for uncertainty quantification for problems of the form $Z=(mT+z_0)^{p}$, where $Z$ is the intensity ratio and $T$ the quantity of interest, is included. 
\end{abstract}

\section{Introduction}

There has been a substantial increase in the availability of remote sensing datasets over the last several decades. This has lead to sustained and intensifying interest in proper statistical modeling of these data for the purposes of retrieval of geophysical fields and parameters, uncertainty quantification, prediction, and other applications.
The fundamental unit of data in many of these applications is a record of the number of photons received by an instrument at a given time, location, and wavelength. These data are further processed into intensity maps, images, and other products. Many applications, notably the algorithm used by the NASA Global Observation of the Limb and Disk (GOLD) mission and others for the retrieval of atmospheric composition ratios \cite{goldon2, icon_on2,strickland_on2,timed_guvi_on2}, some proposed algorithms for the retrieval of thermospheric neutral temperature via far-ultraviolet remote sensing \cite{cantrallthesis, cantrallmatsuo,mine_permproc,zhangbandratio}, or algorithms used for the calculation of hardness ratios and molecular compositions in x-ray astronomy \cite{boersma2012spatial,hardnessratio, Park_2006, wang2024analysis} are based on the idea that ratios of data in specific sub-bands can be related through a simple forward model to parameters and fields of interest. Other applications can be found in  \cite{clayton_ratioassimilation,biasisotoperatios,genbp_sar,horanyi2025interstellar,Jia:14,telikicherla2025improving,f2height_ratio} and others. 
Our approach, influenced by \cite{hardnessratio} and \cite{Park_2006}, is based on the idea that it is the ratio of Poisson means, not the ratio of counts, that is the fundamental physical variable of interest. 
Since the Poisson mean is a latent random variable, many of the applications listed above must be approached from the standpoint of hierarchical Bayesian modeling where we first infer the distribution of the Poisson means, then the distribution of the ratio, and then finally the distribution of the quantity of interest, for example temperature or composition. 

In Section \ref{sec:permprocmodel} we discuss the model implemented in the package, which yields closed-form posterior distributions for the quantity of interest under certain restrictions on the forward model. The model is based on a type of point process model known as a permanental process \cite{mccullagh_permproc,moller2005properties}. Then in Section \ref{sec:demo} is a brief demonstration of the method on some toy problems, and in Section \ref{sec:package} 
we will discuss the specifics of the \texttt{PoissonRatioUQ} package, which is implemented in \rlang \cite{R_citation}. The last two sections contain a brief discussion of future work and a short conclusion. The package is available through github at \texttt{https://github.com/mfleduc/PoissonRatioUQ}, and the first release is available on Zenodo \cite{leduc_poissonratiouq_2026}.

\section{The Model}\label{sec:permprocmodel}

This section gives an overview of the model and estimation techniques. The package also implements an estimation technique that does the estimation of $Z$ in a pointwise manner neglecting spatial correlation, the algorithm is similar to that in \cite{hardnessratio} and \cite{Park_2006}, and is described in \cite{mine_permproc}.

\subsection{The Permanental Process}

At the core of the algorithm is a type of Poisson point process (PPP, \cite{streit.ppp}) known as the permanental process. The permanental process is a PPP with a random intensity function $\lambda(s) = \frac{1}{2}\left(f_1(s)^2+...+f_2(s)^2\right)$ with each $f_i(s) \sim \mathcal{GP}(0,k(s_1,s_2))$. Its name comes from the fact that its nth order product densities are given by \cite{mccullagh_permproc}

\begin{equation}
    \rho^{(n)}(s_1,...,s_n)=\Expect{\prod_{i=1}^n\lambda(s_i)} = \text{per}_{\alpha}(\mathbf{K})
\end{equation}
where $\mathbf{K}_{ij}=k(s_i,s_j)$, $\alpha=n/2$, and $\text{per}_{\alpha}$ is the $\alpha$-weighted permanent given by 

\[
\text{per}_{\alpha}(\mathbf{K})=\sum_{\sigma(1,...,n)}\alpha^{\#\sigma}\prod_{i=1}^nK(s_i,s_{\sigma(i)})
\]
where $\sigma(1,...,n)$ are the permutations of $1,...,n$ and $\#\sigma$ is the number of cycles in the permutation. 

An important property of these processes is that they exhibit point clustering: for a permanental process the pair correlation function, which measures the attraction between two events, is given by \cite{moller2005properties}

\begin{equation}
    \rho^{(2)}(s_1,s_2) = 1 + \frac{k(s_1,s_2)^2}{\alpha \lambda(s_1)\lambda(s_2)}
\end{equation}
making them useful for processes that exhibit these behaviors, such as modeling the distribution of a species of tree in a given area as done in \cite{pmlr-v54-flaxman17a}. These have become useful models in recent years due to their flexibility and ease of estimation via the Representer Theorem \cite{nashed_and_wahba}. 
The next section, which builds off the results of \cite{pmlr-v54-flaxman17a} and \cite{pmlr-v70-walder17a}, describes an estimation scheme for ratios of Poisson intensities based on the permanental process, which was described in \cite{mine_permproc}. 
Due to identifiability issues, the procedure focuses on the case $n=1$.

\subsection{Estimation}

This section presents the method described in \cite{mine_permproc}, which is based off the method implemented in \cite{pmlr-v54-flaxman17a} and \cite{pmlr-v70-walder17a} but assumes that the available data are spatially binned. Estimation for the case where the data is not spatially binned, while mathematically very similar, has not been done and is planned for the future. 

Consider a process on a spatial region $S$ presented as counts observed in $d$ disjoint binds $R_i$, with the number of observed counts in each region $a_i$ distributed as $\Pois{\Lambda_i}$. In this case, the log-likelihood of the intensity vector $\vec{\Lambda} = \left[\Lambda_1, ...,\Lambda_d\right]^T$ is given by

\begin{equation}\label{eq:loglike}
         \ell(\vec{\Lambda}|\{a_i\}_{i=1}^d)  = \sum_{i=1}^da_i \log(\Lambda_i) - \sum_{i=1}^d\Lambda_i  
\end{equation}
Now say that $\vec{\Lambda}$ is driven by a latent vector $\vec{f}\sim \mathcal{N}(0, \gamma^{-1}\mathbf{K})$, such that $\Lambda_i=\frac{c}{2}f_i^2$. Under this model the log-posterior for the latent vector $\vec{f}$ is given by

\begin{equation}
     \ell(\vec{f}|\{a_i\}_{i=1}^d) = \sum_{i=1}^da_i\log\left(\frac{c}{2}f_i^2\right) - \frac{c}{2}\left<\vec{f},\vec{f}\right>-\frac{\gamma}{2}\left<\vec{f},\mathbf{K}^{-1}\vec{f}\right>
\end{equation}
Combining the inner products allows us to write the log-posterior as
\begin{equation}
    \label{eq:logpost}
     \ell(\vec{f}|\{a_i\}_{i=1}^d) = \sum_{i=1}^da_i\log\left(\frac{c}{2}f_i^2\right) - \frac{1}{2}\left<\vec{f},\tilde{\mathbf{K}}^{-1}\vec{f}\right>
\end{equation}
where $\tilde{\mathbf{K}}^{-1} = cI+\gamma\mathbf{K}^{-1}$. The equivalent kernel matrix $\tilde{\mathbf{K}}$ can be calculated from the eigendecomposition of $\mathbf{K}$ as 
\begin{equation}\label{eq:equivalent_kernel}
\begin{split}
    \tilde{\mathbf{K}} =& \mathbf{\Phi}\tilde{\mathbf{H}}\mathbf{\Phi}^T, \\
    \mathbf{\tilde{H}} =& \mathrm{diag}\left(\frac{\eta_j}{c\eta_j+\gamma}\right)   \end{split}
\end{equation}
where $\mathbf{\Phi}_{ij}$ is the value of the $j^{th}$ eigenvector of $\mathbf{K}$ at location $i$ and $\mathbf{H}$ is a diagonal matrix with the eigenvalues $\eta_j$ of $\mathbf{K}$ along the diagonal.

Applying the Representer Theorem \cite{nashed_and_wahba}, we can say that $\vec{f}=\tilde{\mathbf{K}}\vec{\psi}$, which allows us to write the log-posterior as
\begin{equation}
    \label{eq:logpost_alpha}
     \ell(\vec{\psi}|\{a_i\}_{i=1}^d) = \sum_{i=1}^da_i\log\left(\frac{c}{2}\left(\tilde{\mathbf{K}}\vec{\psi}\right)_i^2\right) - \frac{1}{2}\left<\vec{\psi},\tilde{\mathbf{K}}^{-1}\vec{\psi}\right>
\end{equation}
In the package, optimization is done using a conjugate gradient method via the \texttt{optim()} function in base \rlang. The optimization uses an analytic gradient, which can be shown to be
\begin{equation}
    \label{eq:gradient}
    \vec{\nabla}_{\vec{\psi}}\ell(\vec{\psi}|\{a_i\}_{i=1}^d) = -\tilde{\mathbf{K}}\vec{\psi}+2\frac{\tilde{\mathbf{K}}\vec{a}}{\tilde{\mathbf{K}}\vec{\psi}}
\end{equation}
where $\vec{a}$ is a vector of the count data and division is done elementwise.
Then the MAP estimator of $\vec{f}$ is given by $\hat{\vec{f}} = \tilde{\mathbf{K}}\hat{\vec{\psi}}$ where $\hat{\vec{\psi}}$ maximizes Eq. \eqref{eq:logpost_alpha}. Using this and $ \hat{\mathbf{\Sigma}}$ the inverse Hessian of the posterior at the MAP, which is given by 

\begin{equation}
    \label{eq:Hessian_at_MAP_D_binned}
    \begin{split}
    \hat{\mathbf{\Sigma}} & = \tilde{\mathbf{K}}-\tilde{\mathbf{K}}\left(\mathbf{D}^{-1}+\tilde{\mathbf{K}}\right)^{-1}\tilde{\mathbf{K}}^T,\\
    \mathbf{D} &= \text{diag}\left(\frac{\hat{\psi}_i^2}{2a_i}\right),
    \end{split}
\end{equation}
we can approximate the posterior distribution of $\vec{f}$ using a Laplace Approximation, which is a normal distribution with the mean given by $\hat{\vec{f}}$ and covariance matrix given by Eq. \eqref{eq:Hessian_at_MAP_D_binned}  \cite{gpml}.
Then the distribution of $\hat{\Lambda}_i = \frac{c}{2}\hat{f}_i^2$ is well-approximated by 
\begin{equation}
\begin{split}
        \hat{\Lambda}_i  &\sim\Gamma(\hat{\alpha}_i,\hat{\beta}_i), \\  
        \hat{\alpha}_i &= \frac{\left(\mu_i^2+\sigma_i^2\right)^2}{2\sigma_i^2\left(2\mu_i^2+\sigma_i^2\right)}\\ \quad
        \hat{\beta}_i &= \frac{\mu_i^2+\sigma_i^2}{\sigma_i^2 c\left(2\mu_i2+\sigma_i^2\right) }
        \end{split}
\end{equation}
where $\mu_i$ and $\sigma_i$ are the posterior mean and standard deviation of $\hat{f}_i$ and $\hat{\alpha}_i$ and $\hat{\beta}_i$ are the shape and rate parameters of the gamma distribution. Note the similarity to \cite{pmlr-v70-walder17a} Sect. 4.1.5, which uses the scale parameter instead of the rate parameter.
The algorithm in this case is effectively a non-parametric Poisson regression with a quadratic link function rather than the usual exponential link function. 

After the estimation is done for both datasets, we can determine the distribution of the ratio $Z_i = \Lambda_{a,i}/\Lambda_{b,i}$, which is a generalized Beta Prime distribution 

\begin{equation}\label{eq:Z_s_distribution}
    Z = \frac{\hat{\Lambda}_{a,i}}{\hat{\Lambda}_{b,i}} \sim BP\left(\hat{\alpha}_{a,i},\hat{\alpha}_{b,i},1, q_i\right),
\end{equation}
where $q_i=\hat{\beta}_{b,i}/\hat{\beta}_{a,i}$. The PDF of the $BP(\alpha, \beta, p, q)$ distribution is given by
\begin{equation}\label{eq:gbppdf}
     p_Z(z)  = \frac{p}{q^{\alpha p}B(\alpha,\beta)} \frac{z^{\alpha p-1}}{(1+(z/q)^p)^{\alpha+\beta}}
\end{equation}
where $B(\alpha,\beta)$ is the Beta function. Now if $Z_i\approx \left(mT_i+z_0\right)^p$,

\begin{equation}
    \hat{T}_i\sim -\frac{z_0}{m}+BP\left(\hat{\alpha}_{a,i},\hat{\alpha}_{b,i},p,\frac{q^{1/p}}{m}\right)
\end{equation}

While doing the estimation in this manner is not strictly necessary, it is done to allow a single data processing pipeline to handle raw point process data as well, which is expected to 
simplify implementation in the future. 

\section{Demonstration}\label{sec:demo}

This section contains short demonstrations of the method on some toy problems. For application to an FUV remote sensing problem please see \cite{mine_permproc}, where the permanental process model is used to estimate thermospheric neutral temperature, which is critical for understanding the effects of drag on satellites in low-Earth orbit \cite{mehta2023satellite, doornbos2006modelling,leonardetal,zesta2016satellitedrag}. This section demonstrates the model on estimating the intensity ratio in Section \ref{sec:binneddata} and on a problem with a nonlinear transformation from the intensity ratio to the quantity of interest (Section \ref{Sec:nonlin_xform}). The first section also includes a brief timing study.

\subsection{Ratio Estimation} \label{sec:binneddata}

For the first test, we apply the permanental process model to the problem of estimating the function 

\begin{equation}\label{eq:timingstudy_eqn}
    \begin{split}
        Z(x) &= \frac{25\sin\left(\frac{\pi}{2}x\right)^2+10}{8\cos\left(\frac{\pi}{2}x\right)^2+10}, \ -1\le x\le 1
    \end{split}
\end{equation}
given binned count data. In this toy problem, the interval $[-1,1]$ is divided into $n$ regions, with $n$ between 10 and 1000, and the Poisson mean in each bin is taken to be the value of the numerator or denominator at the bin center, depending on which process generates the data.

\begin{figure}[h]
     \centering
     \begin{subfigure}[t]{0.45\textwidth}
         \centering
         \includegraphics[width=\textwidth]{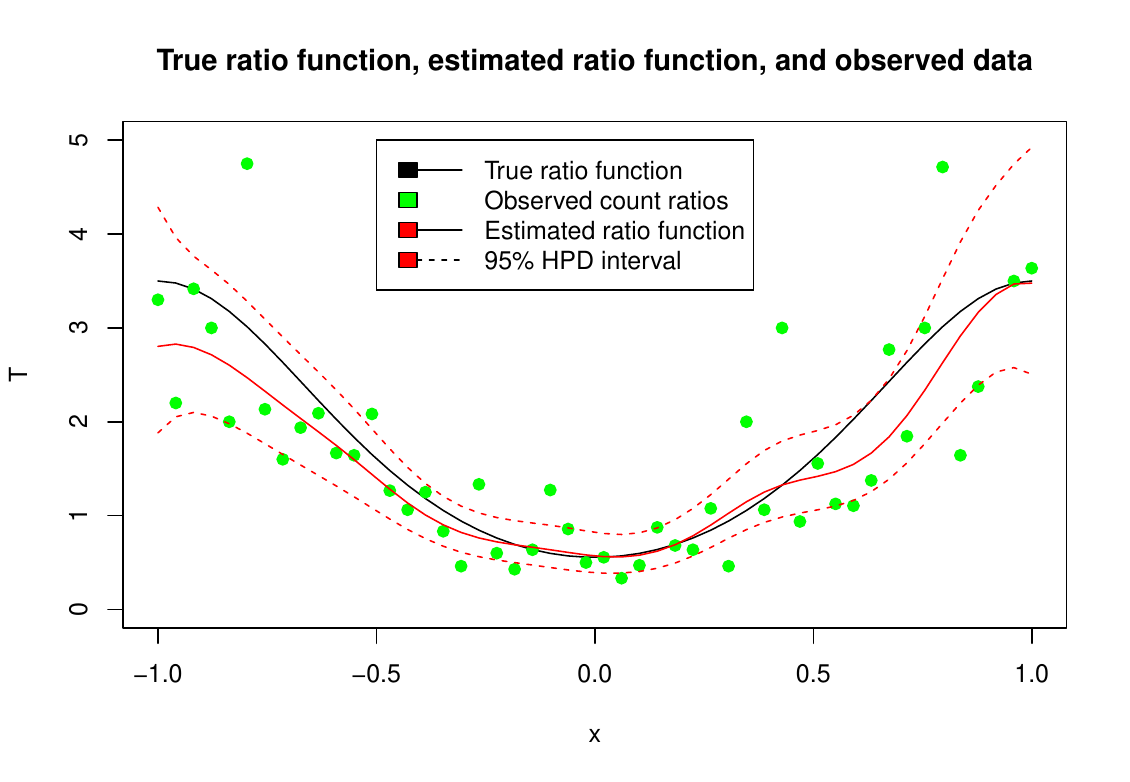}
         \caption{True (black) vs estimated (red) ratio function along with the ratios of observed counts at each bin (green) with 50 bins. The edges of the 95\% HPD interval of the estimate are shown as dotted lines.}
         \label{fig:true_v_est}
     \end{subfigure}
     \hfill
     \begin{subfigure}[t]{0.45\textwidth}
         \centering
         \includegraphics[width=\textwidth]{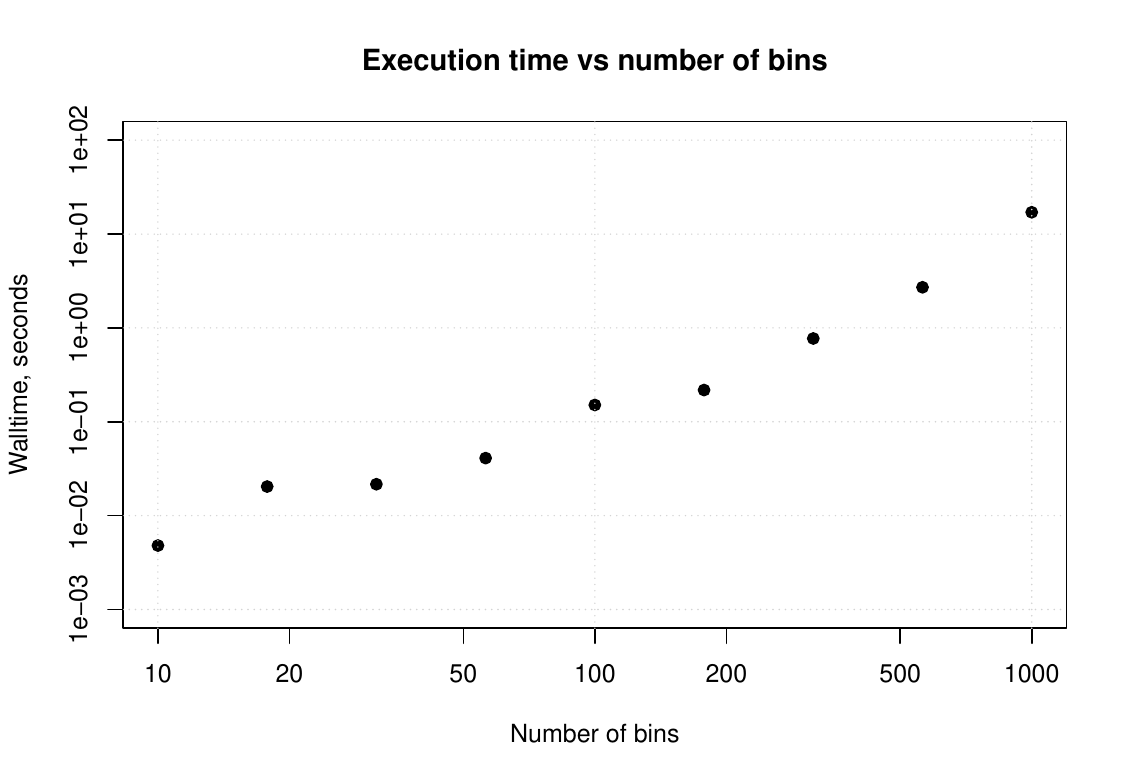}
         \caption{Execution time vs number of bins for bin numbers logarithmically spaced from 10 to 1000. Walltime is a mean over 5 trials.}
         \label{fig:timing_study}
     \end{subfigure}
     \caption{Example retrieval and timing study using the permanental process model to retrieve the ratio function in Eq. \eqref{eq:timingstudy_eqn}}
     \label{fig:timing_and_estimation}
\end{figure}

The results of the estimation procedure are shown in Figure \ref{fig:timing_and_estimation}. In Figure \ref{fig:true_v_est} is a plot of the true vs estimated ratio, along with the ratios of the observed counts, along with the edges of the 95\% highest posterior density set, defined as the set $I_{0.95}$ such that, if $p(z)$ is the posterior density, \cite{Amaral_Turkman_Paulino_Muller_2019}

\[
\int_{I_{0.95}}p(z)\diffd z = 0.95 
\]
where, $\forall x\in I_{0.95}, y\in I_{0.95}^C, \ p(x)\ge p(y)$. The $\alpha-$HPD set refers to the set $I_{\alpha}$ defined in this manner.

Estimation was done using a Wendland kernel with support width of $0.75$ \cite{wendlandrbf}. The estimated curve tracks the true curve fairly closely as expected. Over the 5 trials with 50 bins, the model achieved a mean Continuous Ranked Probability Score (CRPS, \cite{scoring}) of approximately 0.12 and the relative MAE of the MAP estimate was about 7\%. In addition, the timing study shows that even with 1000 bins and a full eigendecomposition of the prior covariance kernel the full posterior distribution of $Z(x)$ is determined in approximately 15 seconds on a desktop computer with 16 GB RAM. 


\subsection{Nonlinear Transformations}\label{Sec:nonlin_xform}

The second test is to demonstrate the model's performance on data with a forward model of the form $Z=(mT+z_0)^p$. Using the same ratio function as in the previous section (Eq. \eqref{eq:timingstudy_eqn}), for the test we are estimating the quantity of interest $T$ given by

\begin{equation}\label{eq:qoi_nonlin}
    T(x) = 5\left(Z(x)^2+2\right)
\end{equation}
, so $m=1/5$, $z_0=-2$, and $p=1/2$.
Under the model $Z(x)\sim BP(\alpha,\beta, 1, q)$, $T(x)$ has a shifted generalized Beta Prime distribution given by

\begin{equation}\label{eq:qoi_dist}
    T(x)\sim -\frac{z_0}{m}+BP\left(\alpha,\beta,p,\frac{q^{1/p}}{m}\right)
\end{equation}

\begin{figure}[h]
     \centering
     \begin{subfigure}[t]{0.49\textwidth}
         \centering
         \includegraphics[width=\textwidth]{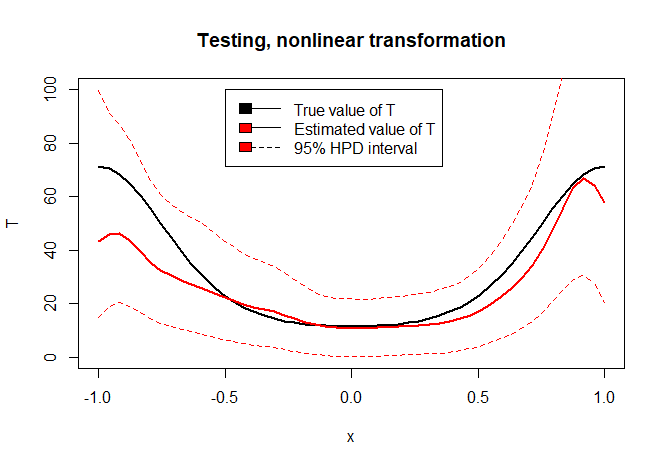}
     \end{subfigure}
     \hfill
     \begin{subfigure}[t]{0.49\textwidth}
         \centering
         \includegraphics[width=\textwidth]{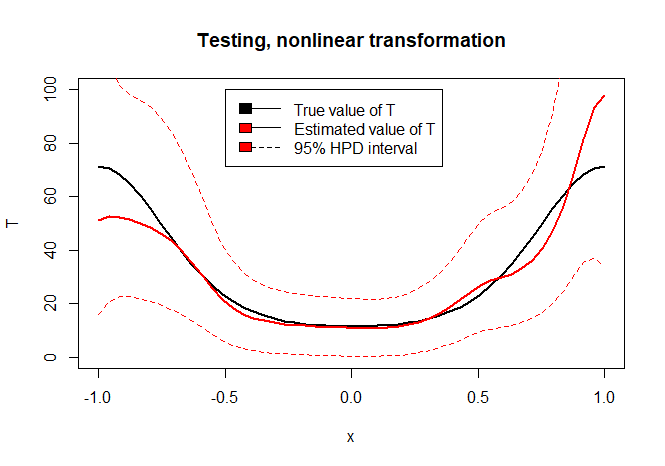}
     \end{subfigure}
     \caption{Estimation results from two realizations of the Poisson data for the nonlinear transformation between the ratio and quantity of interest given by Eq. \eqref{eq:qoi_nonlin} along with estimated 95\% HPD intervals.}
     \label{fig:nonlin_res}
\end{figure}

The estimated value of $T(x)$ along with 95\% HPD intervals are shown in Figure \ref{fig:nonlin_res}. The estimates are fairly accurate in both cases, although the nonlinearity in the transformation can amplify errors in the estimation of $Z$. When $Z>1$ and $p>1$, any errors in estimation are made larger by the nonlinearity, and likewise when both are less than 1. This also results in the extremely wide HPD intervals near the endpoints of the domain. 

\section{The \texttt{PoissonRatioUQ} Package}\label{sec:package}

The package can be installed from github using the command
\begin{verbatim}
    >devtools::install_github('mfleduc/PoissonRatioUQ')
\end{verbatim}
Any issues with the package can be brought to my attention via email or through github. 
What follows is a brief description of the functions contained in the package. 

\subsection{ Functions for Estimation and Calculations with the Generalized Beta Prime Distribution}

This subsection goes over the basic functions contained in the package used for estimation and basic calculations for the generalized Beta Prime distribution. This includes operations on the distribution as well as the main functions used to determine posterior distributions of intensity ratios and quantites of interest. 

\subsubsection{Basic calculations with the generalized Beta Prime distribution}

The package contains four functions to allow for basic calculations with the generalized Beta Prime distribution: \texttt{dbetaprime(), pbetaprime(), qbetaprime(),} and \texttt{rbetaprime()}. The first two calculate the generalized Beta Prime PDF and CDF respectively, the third the quantile function, and the last one generates random numbers with a generalized Beta Prime distribution. 

The PDF of the generalized Beta Prime distribution is given by Eq. \eqref{eq:gbppdf}. Due to numerical stability concerns, the calculation is done in log-space before being converted to the form in Eq. \eqref{eq:gbppdf} at the output stage. 

The calculation of the CDF is given by a known relationship between the Beta Prime and Beta distributions. First, say that $X\sim BP(\alpha, \beta, p, q)$. Then it can be shown that 

\[
Y = \left(\frac{X}{q}\right)^p \sim BP(\alpha, \beta)
\]
and that 

\[
\frac{Y}{1+Y}\sim Beta(\alpha, \beta)
\]
Combining these observations, we can see that the CDF of $X$ is given by

\begin{equation}
    \label{eq:gbp_cdf}
    \begin{split}
    F_X(x) &= I_{z}(\alpha, \beta)\\
    z &= \frac{x^p}{q^p+x^p}
    \end{split}
\end{equation}
where $I_z(\alpha, \beta)$ is the regularized incomplete Beta function (\cite{NIST:DLMF} Eq. 8.17.2). From this, the value of the CDF at $x$ can be easily calculated using the \texttt{pbeta()} function in base \rlang. This process is inverted to calculate the quantile function in \texttt{qbetaprime()} using the \texttt{qbeta()} function. 

\begin{figure}
     \centering
     \begin{subfigure}[b]{0.75\textwidth}
         \centering
         \includegraphics[width=\textwidth]{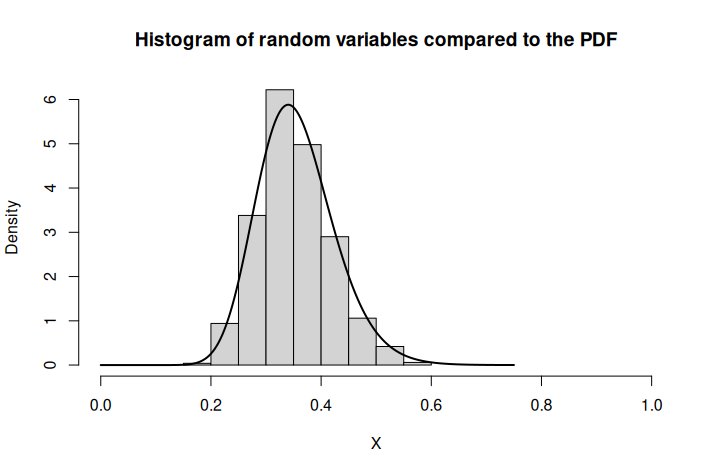}
         \caption{Comparison: Histogram of random variables from \texttt{rbetaprime()} vs analytic PDF}
         \label{fig:rand_v_pdf}
     \end{subfigure}
     \hfill
     \begin{subfigure}[b]{0.49 \textwidth}
         \centering
         \includegraphics[width=\textwidth]{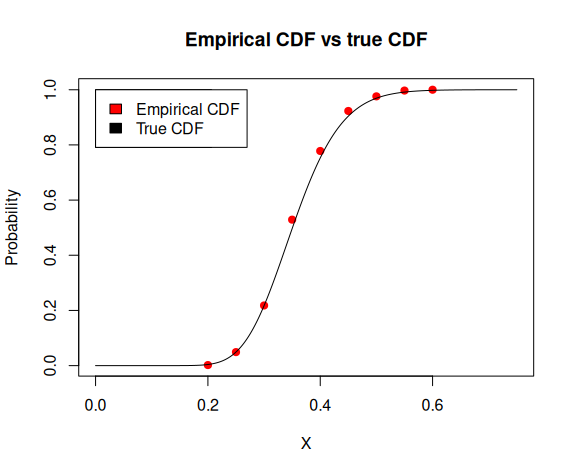}
         \caption{Comparison: Empirical CDF of random variables from \texttt{rbetaprime()} vs analytic CDF}
         \label{fig:rand_v_cdf}
     \end{subfigure}
     \hfill
     \begin{subfigure}[b]{0.49 \textwidth}
         \centering
         \includegraphics[width=\textwidth]{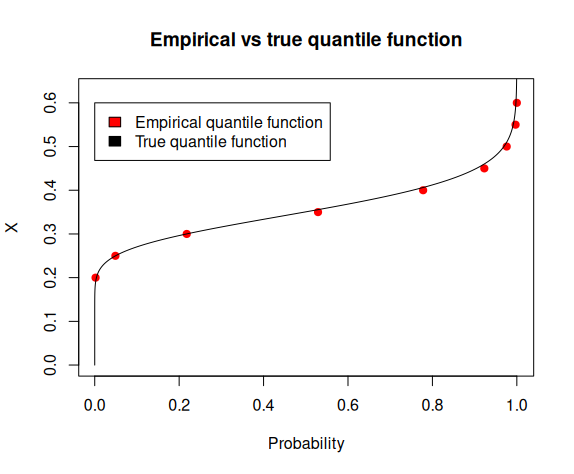}
         \caption{Comparison: Empirical quantiles of random variables from \texttt{rbetaprime()} vs analytic quantiles}
         \label{fig:rand_v_quant}
     \end{subfigure}
        \caption{Comparison: Empirical distribution, CDF, and quantiles from \texttt{rbetaprime()} compared to analytic expressions. Plots generated with 1000 random draws from the distribution.}
        \label{fig:rand_v_analytic}
\end{figure}

This same relationship enables efficient simulation of random numbers with a $BP(\alpha, \beta, p,q)$ distribution. By taking $\{Z_i\}_{i=1}^n\overset{iid}{\sim}Beta(\alpha, \beta)$, the random variables given by

\[
X_i = q\left(\frac{Z_i}{1-Z_i}\right)^{1/p}
\]
are IID with a $BP(\alpha,\beta, p, q)$ distribution. This is the method implemented for random variable generation in \texttt{rbetaprime()}. These functions are all compared to each other in Figure \ref{fig:rand_v_analytic} and show the expected very close agreement between empirical and analytic expressions. The code used to generate the curves shown in the plots is shown below.

\begin{verbatim}
> alpha <- 10;beta <- 20;p <- 2; q <- 0.5 #Parameters of gen. BP
> nrvs <- 1000 #Number of random draws
> X <- rbetaprime( nrvs, alpha, beta, p=p, q=q )
> ecdf <- hist(X, plot=FALSE)
> pdf <- dbetaprime(seq(0,0.75, by=0.0001), alpha, beta, p=p, q=q )
> cdf <- pbetaprime(seq(0,0.75, by=0.0001), alpha, beta, p=p, q=q )
> qtile <- qbetaprime(seq(0, 1, by=0.003), alpha, beta, p=p, q=q )
\end{verbatim}

\subsubsection{Functions for Permanental Process Estimation}

The fundamental building block of the package is the function \texttt{permprocest()}. This function takes as inputs the kernel matrix $\mathbf{K}$, the count data, and the parameters $g$ and $c$, and returns the MAP estimate of the intensity vector $\vec{\Lambda}$ along with the parameters of the posterior distribution of $\Lambda_i$ at each bin and the coefficients of the equivalent kernel expression $\hat{\vec{\psi}}$. The function is called as

\begin{verbatim}
> results <- permprocest(K, counts=count_data, g=g, c=c, maxiter=maxiter)    
\end{verbatim}
where \texttt{K} is the kernel matrix, \texttt{count\_data} is a vector of observed counts in each bin (or a matrix with columns corresponding to realizations of the process), \texttt{g} and \texttt{c} are the $\gamma$ (marginal precision) and $c$ (scaling) parameters, which both default to 1, and \texttt{maxiter} is the maximum number of iterations to perform in the estimation, which defaults to 300. There are functions implemented for calculating both the objective and gradient of the objective for input to the optimization, however those are not discussed here. 

The estimation of the intensity ratio is handled by the function \texttt{ratioestimationpermproc()}. This function takes as minimal inputs a kernel matrix \texttt{K} and count data for the numerator and denominator of the model, and returns the MAP estimate of the intensity ratio as well as the parameters of the generalized Beta Prime distribution describing the posterior. The function also allows the user to input a different kernel matrix for each process, as well as different $\gamma$ and $c$ parameters. The function can be called by 
\begin{verbatim}
> results <- ratioestimationpermproc( K_num, counts_num, counts_denom,
             K2 = K_denom, c1=c1, c2=c2, g1=g1, g2=g2, maxiter=maxiter  )    
\end{verbatim}
where \texttt{K2} is the kernel matrix for the denominator process, and \texttt{c1,c2} and \texttt{g1, g2} are the $c$ and $\gamma$ parameters for the numerator and denominator processes. 

The last function in this section is \texttt{zbetaprime()}. This function implements a slight generalization to the method of \cite{hardnessratio} for estimating the intensity ratio. The algorithm has no spatial component and assumes that locations are independent. It implements a simple Bayesian update to the estimation of the Poisson intensity in each channel using a conjugate Gamma distribution prior, which can be specified, to determine the posterior distribution of the intensity ratio at each location. Because of this updating it can easily handle spatial locations with different numbers of realizations of the process, for example when doing an extra binning step to increase SNR. If this is the case, the count data should be in an array with a new row for each location and a new column for each realization. If a location has no data for a particular realization, it should instead have a value of NaN. The function is called as 
\begin{verbatim}
> results <- zbetaprime(counts_num, counts_denom, a1=a1, b1=b1, a2=a2, b2=b2)
\end{verbatim}
where \texttt{a1,b1,a2,b2} are the shape and rate parameters of the Gamma prior for the numerator and denominator processes. By default the prior distributions are flat, equivalent to shape parameters of 1 and rate parameters of 0. The function returns the MAP estimate of the intensity ratio as well as the parameters of the generalized Beta Prime distribution at each location. Since $p$ does not enter into the estimation of $Z$ at any point, this parameter will always be returned as 1. Further mathematical details are given in \cite{hardnessratio}, as well as in Appendix A of \cite{mine_permproc}.

\subsubsection{ Estimating the Quantity of Interest }

The last function in the package is the function \texttt{tgivenab()}, which is meant to allow estimation of the distribution of the quantity of interest $T$ given the data (Eq. \eqref{eq:qoi_dist}). The function first calls either \texttt{ratioestimationpermproc()} or \texttt{zbetaprime()} depending on the desired model, then calculates the posterior distribution of the quantity of interest given the parameters of the model $Z=(mT+z_0)^p$, which can vary over space. In the event that the modeler does not wish to incorporate spatial correlation in the estimation, the function can be called as
\begin{verbatim}
> results <- tgivenab( counts_num, counts_denom, m,z0,p,spatial=FALSE,
                a1=a1,a2=a2, b1=b1, b2=b2 )
\end{verbatim}
and if the modeler wishes to incorporate spatial structure through the permanental process model, the function can be called as
\begin{verbatim}
> results <- tgivenab( counts_num, counts_denom, m,z0,p,spatial=TRUE,
                K1=K_num, K2=K_denom, c1=c1, c2=c2, g1=g1, g2=g2 )
\end{verbatim}
The variables in each call have the same meaning and defaults as for \texttt{zbetaprime()} and \\ \texttt{ratioestimationpermproc()} respectively. The default behavior is \texttt{spatial=TRUE}.
\subsection{ Functions for Model Scoring and Uncertainty Quantification}

\subsubsection{\texttt{CRPS()}, \texttt{CRPSgaussian()}, and \texttt{CRPSgbp()}}

The package contains three functions that can be used to calculate the CRPS of the predictive model. The first function, \texttt{CRPS()}, is used for an arbitrary distribution given either its probability density function or cumulative density function. The second and third, \texttt{CRPSgaussian()} and \texttt{CRPSgbp()}, are an implementation of the CRPS for the Gaussian and generalized Beta Prime distributions, which both take on a convenient form \cite{scoring,leduc2026continuous}.

The CRPS is given by, for general predictive CDFs $F(y)$ and ground truth value $\hat{x}$, 

\begin{equation}
    \label{eq:CRPS}
    CRPS(F|\hat{x}) = \intR \left(F(y)-H(y-\hat{x})\right)^2\diffd y
\end{equation}
where $H$ is the Heaviside step function. For a Gaussian distribution, this has the more convenient form 
\begin{equation}
    \label{eq:CRPS_Gauss}
    CRPS(F|\hat{x}) = -\sigma\left(\pi^{-1/2}-2\varphi\left(\frac{\hat{x}-\mu}{\sigma}\right) -\frac{\hat{x}-\mu}{\sigma} \left(2\Phi\left(\frac{\hat{x}-\mu}{\sigma}\right)-1\right)\right)
\end{equation}
where $\sigma$ and $\mu$ are the predictive standard deviation and mean, $\varphi()$ is the standard normal PDF, and $\Phi()$ the standard normal CDF \cite{scoring}. The form for the generalized Beta-prime was first shown in \cite{leduc2026continuous} and has the form
 \begin{equation}\label{eq:gbp_crps}
\begin{split}
    &CRPS(F|\hat{x}) = \\
    &2\mu - \hat{x} +\frac{2}{B(\alpha, \beta)}\left(\hat{x}B(w;\alpha,\beta)+\frac{\hat{x}}{\alpha+\beta-1} w^{\alpha-1}(1-w)^\beta\left(1-\tfo{1,\alpha+\beta-1;\alpha+p^{-1},w }\right)  \right)\\
    &-\frac{2q}{\alpha B(\alpha,\beta)^2}B(2\alpha+p^{-1},2\beta-p^{-1}) {}_3F_2(\alpha+\beta, 1, 2\alpha+p^{-1};\alpha+1,2\alpha+2\beta;1)
    \end{split}
\end{equation}
where $B(x;\alpha,\beta)$ is the incomplete Beta function and $w=\frac{\hat{x}^p}{q^p+\hat{x}^p}$. This is more complex but is readily evaluated in \rlang.

The function \texttt{CRPS()} takes three inputs, with the details depending on exactly what is available to the user. The first input is the grid on which the distribution is evaluated, the second is either the PDF or CDF of the predictive distribution, and the last is the value $\hat{x}$. If given the PDF, the function calculates the CDF using the provided grid and a linear interpolation. Then the trapezoid rule is used to calculate the integral in Eq. \eqref{eq:CRPS}. Testing showed that for unimodal parametric distributions, which is the use case this package is meant for, this was sufficiently accurate. 

For Gaussian predictive distributions, or predictive distributions that are sufficiently close to being Gaussian, the function \texttt{CRPSGaussian()} is available as an alternative. This function takes as arguments the mean $\mu$, standard deviation $\sigma$, and ground truth $\hat{x}$ and evaluates the CRPS using Eq. \eqref{eq:CRPS_Gauss}. Similarly the function \texttt{CRPSgbp()} takes the parameters $\alpha,\beta,p,q,$ and $\hat{x}$ as arguments and computes the CRPS using Eq. \eqref{eq:gbp_crps}. The function \texttt{inc\_beta()}, which evaluates the incomplete beta function with parameters $a,b$ at points $w$, is included to assist in calculations. A quick test, whose outcome is shown below, was performed to assess the accuracy and speed of these calculations. 
\begin{verbatim}
> xhat <- 1
### First: Test the Gaussian model
> mu <- 0 
> sigma <- 1.5
> yvals <- seq(-5,5,by=0.001)
> crps_g <- CRPSgaussian(mu,sigma,xhat)
> print(crps_g)
[1] 0.6070746
> npdf <- dnorm( yvals, mean=mu, sd=sigma )
> crps_int <- CRPS(yvals, pdf=npdf,xhat=xhat)
> print(crps_int)
[1] 0.6066229
> ncdf <- pnorm(yvals, mean=mu, sd=sigma)
> crps_cdf <- CRPS(yvals, cdf=ncdf, xhat=xhat)
> print(crps_cdf)
[1] 0.606827
### Now: the GBP
> alpha <- 2
> beta <- 3
> yvals <- seq(0,10,by=0.001
> crps_gbp <- CRPSgbp(alpha,beta,xhat,p=1,q=1)
> print(crps_gbp)
[1] 0.2357143
> pdf_gbp <- dbetaprime(yvals,alpha,beta,p=1,q=1)
> crps_int <- CRPS(yvals,pdf_gbp,xhat)
> print(crps_int)
[1] 0.2348803
\end{verbatim}
All of these calculations show a relative error of less than 0.4\%, which is sufficient for our purposes. Since this package is intended for use with parametric distributions, it can in theory achieve arbitrary accuracy based on the precision of the input grid. However, if the distribution to be evaluated is a Gaussian or generalized Beta-prime, the specific functions should be used as they are exact expressions and much faster than evaluating the integral in Eq. \eqref{eq:CRPS}.

\subsubsection{The functions \texttt{hpdset()} and \texttt{hpdintervalgaussian()}}

The last two functions in the package are \texttt{hpdset()} and \texttt{hpdintervalgaussian()}, which are used to calculate the highest posterior density sets for an arbitrary distribution and the Gaussian distribution respectively. Additional discussion of the highest posterior density set is included in \cite{Amaral_Turkman_Paulino_Muller_2019}. The calculation for a Gaussian distribution is straightforward, requiring only the values of the standard normal quantile function at $\frac{1}{2}(1\pm\alpha)$. For the non-Gaussian case, the calculation requires solving the problem 
\begin{equation}
    \label{eq:hpd_opt_prob}
    \int_{f(x)\ge h}f(x)\diffd x = \alpha
\end{equation}
for $h$. 

The function \texttt{hpdset()} requires three inputs: the evaluation grid, the values of the PDF $f(x)$ at the grid, and the coverage $\alpha$. The calculation proceeds first by checking the normalization of the PDF, then determining the value of $f(x)$ at the mode. The function then uses this value as the initial guess for $h$ in a rootfinding algorithm applied to the function 
\[
g(h) =  \int_{f(x)\ge h}f(x)\diffd x - \alpha
\]
using the \texttt{uniroot()} function in base \rlang. The function then returns all of the values of $x$ from the original input grid that are inside the $\alpha-$HPD set. This function is capable of calculating the HPD set for multimodal distribution functions as well.

The following code shows an example of this method applied to a standard normal distribution, with a comparison between the functions \texttt{hpdset()} and \texttt{hpdintervalgaussian()}. Again the accuracy is mostly constrained by the resolution of the underlying grid. 

\begin{verbatim}
> mu <- 0
> sigma <- 1
> alpha <- 0.95 #Desired coverage ratio
> #Gaussian HPD
> hpd_g <- hpdintervalgaussian( mu, sigma, alpha )
> print(hpd_g[1]) #Endpoints of the Gaussian HPD using the quantile function
[1] -1.959964
> print(hpd_g[length(hpd_g)])
[1] 1.959964
> # Using the integral formulation 
> yvals <- seq(-5,5,by=0.001)
> pdf <- dnorm(yvals, mean=mu,sd=sigma)
> hpd_int <- hpdset( yvals, pdf, alpha  )
> print(hpd_int[1]) #Endpoints of the HPD calculated using the integral formulation 
[1] -1.959
> print(hpd_int[length(hpd_int)])
[1] 1.959
\end{verbatim}

\begin{figure}
    \centering
    \includegraphics[width=0.85\linewidth]{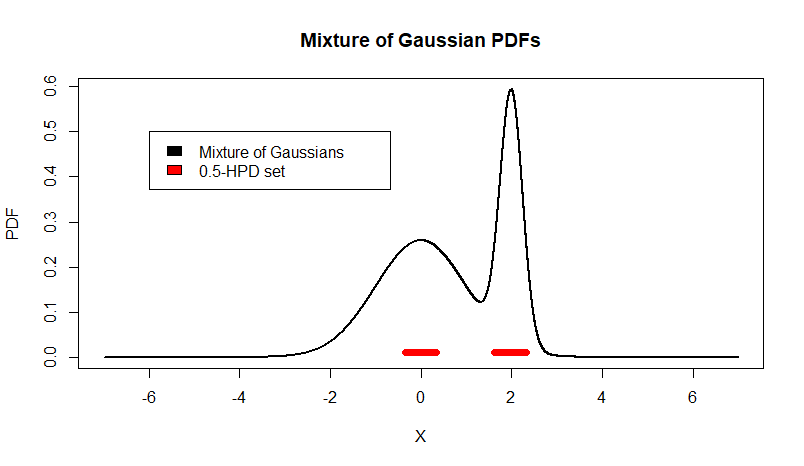}
    \caption{Demonstration of the algorithm for calculating the highest-posterior density sets on a mixture of Gaussians. The algorithm is able to capture the bimodal nature of the distribution. }
    \label{fig:hpd_mjultimodal}
\end{figure}

As a second test, I have calculated the 0.5-HPD set for a mixture of Gaussians given by 

\[
p_X(x) = 0.65\mathcal{N}(\mu=0,\sigma=1) + 0.35\mathcal{N}(\mu=2, \sigma=0.25). 
\]
The code used to do the calculation is reported below, and the PDF along with the calculated HPD set is shown in Figure \ref{fig:hpd_mjultimodal}. The algorithm accurately captures the multimodal nature of the PDF.

\begin{verbatim}
mu <- 0
sigma <- 1
mu2 <- 2
sigma2 <- 0.25
alpha <- 0.5
wght1 <- 0.65
wght2 <- 1-wght1
# Using the integral formulation 
yhat <- seq(-7,7,by=0.001)
pdf <- dnorm(yhat, mean=mu,sd=sigma)
pdf <- wght1*pdf + wght2*dnorm( yhat, mean=mu2, sd=sigma2 )
hpd_int <- hpdset( yhat, pdf, alpha  )
\end{verbatim}

\section{Future Work}\label{sec:future}

Two immediate applications for future work invole the ability to exploit sparsity in the equivalent kernel matrix to speed up model fitting, as well as allowing low-rank decompositions of the kernel matrix or a precalculated eigendecomposition to prevent needing to calculate the entire eigendecomposition. This is especially important for large problems, as the decomposition is $O(n^3)$ and thus begins to dominate the computation, which is $O(n^2)$ otherwise. 

Other than some performance improvements that will allow the package to be used on larger datasets, the biggest goal for the future is the implementation of the method for unbinned data. In that case, the goal is to estimate $f(s)\sim \mathcal{GP}(0, \gamma^{-1}k(s_1,s_2)$ based on observed locations where an event (an emission, for example) occurred. The math works out similarly, with
the log-posterior distribution as, letting $\lambda(s)=\frac{c}{2}f(s)^2$,

\begin{equation}\label{eq:logpost_rkhs}
    \ell(f|\{s_i\}_{i=1}^d) = \sum_{i=1}^d \log\left(\frac{c}{2}f(s_i)^2\right) - \frac{c}{2}\|f\|_2^2 - \frac{\gamma}{2}\|f\|_{\mathcal{H}_k}^2
\end{equation}
where $\|f\|_k$ is the norm in the Reproducing Hilbert Space generated by the kernel $k(s_1,s_2)$, i.e the set 

\begin{equation}
    \label{eq:rkhs_def}
    \mathcal{H}_k(S)  = \{f\in \mathcal{L}^2(S): f(s) = \int_Sk(s,x)g(x)\diffd \mu(x), g(x)\in \mathcal{L}^2(S)\}
\end{equation}
 Since $k(s_1,s_2)$ is a positive definite kernel, it has a Mercer expansion $k(s_1,s_2)=\sum\eta_j\phi_j(s_1)\phi_j(s_2)$, where $\{\phi_j\}_{j=1}^{\infty}$ is an orthonormal basis of $\mathcal{L}^2(S)$ \cite{mercer_representation}.

The major issue is due to the fact that the observation locations are randomly distributed and that $\lambda(s)$ is actually the Radon derivative of the Poisson intensity (i.e the number of events in a set $R$ is Poisson distributed with mean $\int_R \lambda(s)\diffd \mu(s)$). Since we do not assume an explicit Mercer expansion of the kernel in the package, we must approximate the eigenvectors and eigenvalues of the operator

\[
T[f](s) = \int_Sk(s,x)f(x)\diffd \mu(x)
\]
via a Nystrom approximation rather than simply using the observation locations (see \cite{pmlr-v54-flaxman17a, pmlr-v70-walder17a} for details). Those are then used to derive an equivalent kernel. This is not yet implemented.

\section{Conclusions}\label{sec:conc}

This report introduces the \texttt{PoissonRatioUQ} package in \rlang, a package developed for the Bayesian estimation of ratios of Poisson intensities and some inverse problems that use these ratios. This report describes the mathematical content of the permanental process model which is implemented in the package. The report also demonstrates the method's performance in retrievals over a range of Poisson intensities and its ability to recover spatially distributed intensity ratios in an efficient manner. The report also describes several auxiliary functions contained in the package that aid in uncertainty quantification and model evaluation. 

\section{Acknowledgements}
This work was partially supported by the NSF AGS-2231409 and NASA-80NSSC22K0175 awards to the University of Colorado Boulder. 

\newpage
\printbibliography[title={}] 

@article{leduc2026continuous,
  title={The Continuous Rank Probability Score of a Generalized Beta-Prime Distribution and Some Special Cases},
  author={LeDuc, Matthew},
  journal={arXiv preprint arXiv:2603.13622},
  year={2026}
}

@article{hardnessratio,
author="Y.K. Jin and S.N. Zhang and J.F. Wu",
title="Hardness 
Ratio estimation in Low Counting X-Ray Photometry",
year=2006,
journal="The Astrophysical Journal",
doi="10.1086/508677"
}

@article{cantrallmatsuo,
title="Deriving column-integrated thermospheric temperature
with the N2 Lyman–Birge–Hopfield (2,0) band",
author="Clayton Cantrall and Tomoko Matsuo",
journal="Atmospheric Measurement Techniques",
doi="10.5194/amt-14-6917-2021",
year=2021}

@phdthesis{cantrallthesis,
    author = "Clayton Cantrall",
    title = "New approaches for quantifying and understanding thermosphere temperature
variability from far ultraviolet dayglow",
    school = "University of Colorado-Boulder",
    year = 2022
}

@article{biasisotoperatios,
title="Statistical Bias in Isotope Ratios",
author="Christopher D. Coath and Robert C. J. Steele and W. Fred Lunnon",
year=2013,
journal=" Journal of Analytical Atomic Spectrometry"}

@article{zhangbandratio,
author = {Zhang, Yongliang and Paxton, Larry J. and Schaefer, Robert K.},
title = {Deriving Thermospheric Temperature From Observations by the Global Ultraviolet Imager on the Thermosphere Ionosphere Mesosphere Energetics and Dynamics Satellite},
journal = {Journal of Geophysical Research: Space Physics},
volume = {124},
number = {7},
pages = {5848-5856},
year = {2019}
}

@article{Park_2006,
doi = {10.1086/507406},
url = {https://dx.doi.org/10.1086/507406},
year = {2006},
month = {nov},
publisher = {},
volume = {652},
number = {1},
pages = {610},
author = {Park, Taeyoung and Kashyap, Vinay L. and Siemiginowska, Aneta and van Dyk, David A. and Zezas, Andreas and Heinke, Craig and Wargelin, Bradford J.},
title = {Bayesian Estimation of Hardness Ratios: Modeling and Computations},
journal = {The Astrophysical Journal}
}

@article{scoring,
author = "T. Gneiting and A. Rafferty",
year=2007,
title="Strictly Proper Scoring Rules, Prediction, and Estimation",
journal="Journal of the American Statistical Association",
url=" https://doi.org/10.1198/016214506000001437"
}

@article{clayton_ratioassimilation,
author = {Cantrall, Clayton E. and Matsuo, Tomoko and Solomon, Stanley C.},
title = {Upper Atmosphere Radiance Data Assimilation: A Feasibility Study for GOLD Far Ultraviolet Observations},
journal = {Journal of Geophysical Research: Space Physics},
volume = {124},
number = {10},
pages = {8154-8164},
url = {https://agupubs.onlinelibrary.wiley.com/doi/abs/10.1029/2019JA026910},
year = {2019}
}

@article{goldon2,
author = {Correira, J. and Evans, J. S. and Lumpe, J. D. and Krywonos, A. and Daniell, R. and Veibell, V. and McClintock, W. E. and Eastes, R. W.},
title = {Thermospheric Composition and Solar EUV Flux From the Global-Scale Observations of the Limb and Disk (GOLD) Mission},
journal = {Journal of Geophysical Research: Space Physics},
volume = {126},
number = {12},
pages = {e2021JA029517},
doi = {https://doi.org/10.1029/2021JA029517},
url = {https://agupubs.onlinelibrary.wiley.com/doi/abs/10.1029/2021JA029517},
year = {2021}
}

@article{Jia:14,
author = {Jingyu Jia and Fan Yi},
journal = {Appl. Opt.},
number = {24},
pages = {5330--5343},
publisher = {Optica Publishing Group},
title = {Atmospheric temperature measurements at altitudes of 5-30km with a double-grating-based pure rotational Raman lidar},
volume = {53},
month = {Aug},
year = {2014},
url = {https://opg.optica.org/ao/abstract.cfm?URI=ao-53-24-5330},
doi = {10.1364/AO.53.005330},
}

@book{streit.ppp,
title="Poisson Point Processes:Imaging, Tracking, and Sensing",
author="Roy L. Streit",
publisher="Springer",
year=2010}

@article{moller2005properties,
  title={Properties of spatial Cox process models},
  author={M{\o}ller, Jesper},
  journal={Journal of Statistical Research of Iran},
  volume={2},
  number={1},
  pages={1--18},
  year={2005},
  publisher={Statistical Research and Training Center}
}

@article{mccullagh_permproc,
author="McCullagh, Peter and Jesper Møller",
title="The Permanental Process",
journal="Advances in Applied Probability",
volume=38, number=4 ,year=2006,pages="873–888",
url="https://doi.org/10.1017/S0001867800001361." }

@article{pmlr-v54-flaxman17a,
  author="Seth Flaxman and Yee Whye Teh and Dino Sejdinovic",
title="Poisson intensity estimation with reproducing kernels", journal="Electronic Journal of Statistics",
volume=11,number=2, pages="5081-5104", year=2017 
}

@article{mercer_representation,
author = {Mercer, James  and Forsyth, Andrew Russell },
title = {XVI. Functions of positive and negative type, and their connection the theory of integral equations},
journal = {Philosophical Transactions of the Royal Society of London. Series A, Containing Papers of a Mathematical or Physical Character},
volume = {209},
number = {441-458},
pages = {415-446},
year = {1909},
doi = {10.1098/rsta.1909.0016},
URL = {https://royalsocietypublishing.org/doi/abs/10.1098/rsta.1909.0016},
eprint = {https://royalsocietypublishing.org/doi/pdf/10.1098/rsta.1909.0016}
}

@InProceedings{pmlr-v70-walder17a,
  title = 	 {Fast {B}ayesian Intensity Estimation for the Permanental Process},
  author =       {Christian J. Walder and Adrian N. Bishop},
  booktitle = 	 {Proceedings of the 34th International Conference on Machine Learning},
  pages = 	 {3579--3588},
  year = 	 {2017},
  editor = 	 {Precup, Doina and Teh, Yee Whye},
  volume = 	 {70},
  series = 	 {Proceedings of Machine Learning Research},
  month = 	 {06--11 Aug},
  publisher =    {PMLR},
  url = 	 {https://proceedings.mlr.press/v70/walder17a.html},
}

@book{gpml,
    author = {Rasmussen, Carl Edward and Williams, Christopher K. I.},
    title = {Gaussian Processes for Machine Learning},
    publisher = {The MIT Press},
    year = {2005},
    month = {11},
    isbn = {9780262256834},
    doi = {10.7551/mitpress/3206.001.0001},
    url = {https://doi.org/10.7551/mitpress/3206.001.0001},
    eprint = {https://direct.mit.edu/book-pdf/2514321/book\_9780262256834.pdf},
}

@article{nashed_and_wahba,
author = "M.Z. Nashed and Grace Wahba",
title="Regularization and Approximation of Linear Operator Equations in Reproducing Kernel Spaces", 
journal="Bulletin of the American Mathematical Society",
year=1974,
volume=80
}

@ARTICLE{genbp_sar,
  author={Gallardo i Peres, Gerard and Dall, Jørgen and Mason, Philippa J. and Ghail, Richard and Hensley, Scott},
  journal={IEEE Transactions on Geoscience and Remote Sensing}, 
  title={A Generalized Beta Prime Distribution as the Ratio Probability Density Function for Change Detection Between Two SAR Intensity Images With Different Number of Looks}, 
  year={2024},
  volume={62},
  number={},
  pages={1-14},
  doi={10.1109/TGRS.2024.3369509}}

@article{boersma2012spatial,
  title={Spatial analysis of the polycyclic aromatic hydrocarbon features southeast of the Orion Bar},
  author={Boersma, C and Rubin, RH and Allamandola, LJ},
  journal={The Astrophysical Journal},
  volume={753},
  number={2},
  pages={168},
  year={2012},
  publisher={IOP Publishing}
}

@article{wang2024analysis,
  title={Analysis of bright source hardness ratios in the 4 yr Insight-HXMT galactic plane scanning survey catalog},
  author={Wang, Chen and Liao, Jin-Yuan and Guan, Ju and Liu, Yuan and Li, Cheng-Kui and Sai, Na and Jin, Jing and Zhang, Shuang-Nan},
  journal={Research in Astronomy and Astrophysics},
  volume={24},
  number={2},
  pages={025013},
  year={2024},
  publisher={IOP Publishing}
}

@Manual{R_citation,
    title = {R: A Language and Environment for Statistical Computing},
    author = {{R Core Team}},
    organization = {R Foundation for Statistical Computing},
    address = {Vienna, Austria},
    year = {2025},
    url = {https://www.R-project.org/},
  }

@misc{mine_permproc,
    author = "Matthew LeDuc and Tomoko Matsuo and William Kleiber",
    title = "A New Approach to Inversion of Multi-spectral Data and
Applications to FUV Remote Sensing",
    url="https://doi.org/10.5194/EGUSPHERE-2025-5570",
year=2025
}

@Article{f2height_ratio,
title = {A new strategy for ionospheric remote sensing using the 130.4/135.6 nm airglow intensity ratios},
journal = {Earth and Planetary Physics},
volume = {7},
number = {4},
pages = {445-459},
year = {2023},
issn = {2096-3955},
doi = {10.26464/epp2023042},	
author = {XiaoHan Yin and JianQi Qin and Larry J. Paxton}
}

@article{wendlandrbf,
author = "Holger Wendland",
title = "Piecewise polynomial, positive definite and compactly supported radial functions of minimal degree.",
journal="Adv Comput Math",
volume=4, 
pages="389–396",
year=1995
}

@book{Amaral_Turkman_Paulino_Muller_2019, place={Cambridge}, series={Institute of Mathematical Statistics Textbooks}, title={Computational Bayesian Statistics: An Introduction}, publisher={Cambridge University Press}, author={Amaral Turkman, M. Antónia and Paulino, Carlos Daniel and Müller, Peter}, year={2019}, collection={Institute of Mathematical Statistics Textbooks}}

@article{timed_guvi_on2,
author = {Zhang, Y. and Paxton, L. J. and Morrison, D. and Wolven, B. and Kil, H. and Meng, C.-I. and Mende, S. B. and Immel, T. J.},
title = {O/N2 changes during 1–4 October 2002 storms: IMAGE SI-13 and TIMED/GUVI observations},
journal = {Journal of Geophysical Research: Space Physics},
volume = {109},
number = {A10},
pages = {},
doi = {https://doi.org/10.1029/2004JA010441},
url = {https://agupubs.onlinelibrary.wiley.com/doi/abs/10.1029/2004JA010441},
eprint = {https://agupubs.onlinelibrary.wiley.com/doi/pdf/10.1029/2004JA010441},
year = {2004}
}

@article{strickland_on2,
author = {Strickland, D. J. and Evans, J. S. and Paxton, L. J.},
title = {Satellite remote sensing of thermospheric O/N2 and solar EUV: 1. Theory},
journal = {Journal of Geophysical Research: Space Physics},
volume = {100},
number = {A7},
pages = {12217-12226},
url = {https://doi.org/10.1029/95JA00574},
year = {1995}
}

@article{icon_on2,
author = {Meier, R. R.},
title = {The Thermospheric Column O/N2 Ratio},
journal = {Journal of Geophysical Research: Space Physics},
volume = {126},
number = {3},
pages = {e2020JA029059},
url = {https://doi.org/10.1029/2020JA029059},
year = {2021}
}

@article{telikicherla2025improving,
  title={Improving Solar Flare Nowcasting with the Hot Onset Precursor Event Technique},
  author={Telikicherla, Anant and Woods, Thomas N and Schwab, Bennet D},
  journal={The Astrophysical Journal},
  volume={993},
  number={1},
  pages={95},
  year={2025},
  publisher={IOP Publishing}
}

@article{horanyi2025interstellar,
  title={Interstellar Dust Experiment (IDEX) onboard NASA’s Interstellar Mapping and Acceleration Probe (IMAP)},
  author={Hor{\'a}nyi, Mih{\'a}ly and Tucker, Scott and Sternovsky, Zoltan and Tyagi, Kush and Knappmiller, Scott and Ayari, Ethan and Mikula, Rebecca and Szalay, Jamey R and Kempf, Sascha and Bollendonk, Chip and others},
  journal={Space Science Reviews},
  volume={221},
  number={8},
  pages={1--41},
  year={2025},
  publisher={Springer}
}

@misc{NIST:DLMF,
         key = "{\relax DLMF}",
       title = "{\it NIST Digital Library of Mathematical Functions}",
howpublished = "\url{https://dlmf.nist.gov/}, Release 1.2.5 of 2025-12-15",
         url = "https://dlmf.nist.gov/",
        note = "F.~W.~J. Olver, A.~B. {Olde Daalhuis}, D.~W. Lozier, B.~I. Schneider,
                R.~F. Boisvert, C.~W. Clark, B.~R. Miller, B.~V. Saunders,
                H.~S. Cohl, and M.~A. McClain, eds."}

@incollection{zesta2016satellitedrag,
  title={Satellite orbital drag},
  author={Zesta, Eftyhia and Huang, Cheryl Y},
  booktitle={Space weather fundamentals},
  pages={329--351},
  year={2016},
  publisher={CRC Press}
}

@article{leonardetal,
year=2012,
author = {Leonard, J. M. and Forbes, J. M. and Born, G. H.},
title = {Impact of tidal density variability on orbital and reentry predictions},
journal = {Space Weather},
volume = {10},
number = {12},
pages = {},
doi = {https://doi.org/10.1029/2012SW000842},
}

@article{doornbos2006modelling,
  title={Modelling of space weather effects on satellite drag},
  author={Doornbos, E and Klinkrad, H},
  journal={Advances in Space Research},
  volume={37},
  number={6},
  pages={1229--1239},
  year={2006},
  publisher={Elsevier}
}

@article{mehta2023satellite,
  title={Satellite drag coefficient modeling for thermosphere science and mission operations},
  author={Mehta, Piyush M and Paul, Smriti N and Crisp, Nicholas H and Sheridan, Philip L and Siemes, Christian and March, G{\"u}nther and Bruinsma, Sean},
  journal={Advances in Space Research},
  volume={72},
  number={12},
  pages={5443--5459},
  year={2023},
  publisher={Elsevier}
}

@software{leduc_poissonratiouq_2026,
  author       = {LeDuc, Matthew},
  title        = {{mfleduc/PoissonRatioUQ: First\_release}},
  month        = jun,
  year         = 2026,
  publisher    = {Zenodo},
  version      = {First\_release},
  doi          = {10.5281/zenodo.20492078}
}
\end{document}